\documentclass[a4paper,11pt]{article}
\pdfoutput=1 

\usepackage{jcappub} 
\usepackage[normalem]{ulem}      
\usepackage{bm}

\usepackage[T1]{fontenc} 
\usepackage{bigints}

\def\lsim{\mathrel{\raise.3ex\hbox{$<$\kern-.75em\lower1ex\hbox{$\sim$}}}}
\def\gsim{\mathrel{\raise.3ex\hbox{$>$\kern-.75em\lower1ex\hbox{$\sim$}}}}

\newcommand{\be}{\begin{equation}}
\newcommand{\ee}{\end{equation}}
\newcommand{\beq}{\begin{equation}}
\newcommand{\eeq}{\end{equation}}
\usepackage{bm}
\newcommand{\de}{{\textrm d}}
\newcommand{\drm}{\textrm}
\usepackage{soul}
\newcommand{\Fermi}{{\sl Fermi}}

\title{\boldmath On the Neutrino and Gamma-Ray Emission from NGC 1068}

\author[a,b]{Carlos Blanco,}
\author[c,d,e]{Dan Hooper,}

\author[b]{Tim Linden,}
\author[c,e]{Elena Pinetti}

\affiliation[a]{Princeton University, Department of Physics, Princeton, NJ 08544, USA}
\affiliation[b]{Stockholm University and the Oskar Klein Centre for Cosmoparticle Physics, Alba Nova, 10691 Stockholm, Sweden}
\affiliation[c]{Fermi National Accelerator Laboratory, Theoretical Astrophysics Department, Batavia, IL 60510, USA}
\affiliation[d]{University of Chicago, Department of Astronomy \& Astrophysics, Chicago, IL 60637, USA}
\affiliation[e]{University of Chicago, Kavli Institute for Cosmological Physics, Chicago, IL 60637, USA}

\emailAdd{carlosblanco2718@princeton.edu, dhooper@fnal.gov, linden@fysik.su.se, epinetti@fnal.gov}

\abstract{IceCube has recently reported the detection of $\sim 1-10 \,{\rm TeV}$ neutrinos from the nearby active galaxy, NGC 1068. The lack of TeV-scale emission from this source suggests that these neutrinos are generated in the dense corona that surrounds NGC 1068's supermassive black hole. In this paper, we present a physical model for this source, including the processes of pair production, pion production, synchrotron, and inverse Compton scattering. We have also performed a new analysis of \Fermi-LAT data from the direction of NGC 1068, finding that the gamma-ray emission from this source is very soft but bright at energies below $\sim 1 \, {\rm GeV}$. Our model can predict a gamma-ray spectrum that is consistent with \Fermi-LAT observations, but only if the magnetic field within the corona of this active galactic nucleus (AGN) is quite high, namely $B\gtrsim 6 \, {\rm kG}$. To explain the observed neutrino emission, this source must accelerate protons with a total power that is comparable to its intrinsic X-ray luminosity. In this context, we consider two additional nearby active galaxies, NGC 4151 and NGC 3079, which have been identified as promising targets for IceCube.}

\begin{document}
\maketitle
\flushbottom

\section{Introduction}

The IceCube Collaboration has reported an excess of 79 events from the nearby active galaxy NGC 1068, corresponding to a 4.2$\sigma$ detection of neutrinos in the $\sim 1-10 \, {\rm TeV}$ energy range~\cite{IceCube:2022der} (see also Ref.~\cite{IceCube:2019cia}). Although this source was detected in the $\sim 0.1-30 \, {\rm GeV}$ band by \Fermi-LAT ~\cite{Fermi-LAT:2019yla,Fermi-LAT:2019pir}, MAGIC has failed to observe gamma rays from NGC 1068, placing strong limits on its emission at higher energies~\cite{MAGIC:2019fvw}. The lack of TeV-scale gamma rays from this source rules out the possibility that the observed neutrinos are produced in an optically thin environment, and instead favors scenarios in which pions are produced in the dense region that immediately surrounds this AGN's supermassive black hole, where gamma rays can be efficiently absorbed through the process of pair production~\cite{Murase:2022dog}.

Several lines of evidence suggest that a significant fraction of the neutrinos detected by IceCube~\cite{IceCube:2020acn,IceCube:2021rpz,IceCube:2013cdw,IceCube:2013low,IceCube:2014stg} may originate from optically thick, or so-called ``hidden'' sources. In particular, if the pion production rate is normalized to the diffuse neutrino spectrum reported by IceCube, one finds that these sources (if optically thin) should collectively generate a flux of gamma rays that would approximately saturate, or even exceed, the isotropic background reported by the \Fermi-LAT Collaboration~\cite{Hooper:2016jls,Murase:2015xka,Giacinti:2015pya,Murase:2013rfa}. Combining this information with the lack of observed correlations between the directions of IceCube's neutrinos and known gamma-ray sources~\cite{IceCube:2023htm,IceCube:2019cia,IceCube:2016tpw,IceCube:2018omy,IceCube:2016ipa,Smith:2020oac,IceCube:2016qvd,Hooper:2018wyk}, transparent source scenarios appear to be somewhat disfavored. In this context, the X-ray dense cores of AGN are a particularly well-motivated class of high-energy neutrino sources~\cite{Murase:2015xka,Khiali:2015tfa,Stecker:2013fxa,Kimura:2014jba,Kalashev:2014vya} (for a review, see Ref.~\cite{Murase:2015ndr}).

In this paper, we describe a simple model for the neutrino and gamma-ray emission observed from NGC 1068. In particular, we use the measured X-ray luminosity of NGC 1068~\cite{Marinucci:2015fqo,Bauer:2014rla} to build a model for the absorption of gamma rays in the AGN's corona, allowing us to easily explain the lack of very high-energy emission from this source. The energetic electrons then undergo synchrotron and inverse Compton scattering, resulting in significant emission at sub-GeV energies. We also perform an analysis of \Fermi-LAT data, finding evidence of a significant gamma-ray flux extending down to energies of at least 70 MeV. To avoid exceeding the gamma-ray emission observed at sub-GeV energies, the vast majority of the energy in the high-energy electrons must be lost to synchrotron, requiring the presence of very large magnetic fields in the AGN's corona, $B \gsim 6 \, {\rm kG}$. Future MeV-scale observations of NGC 1068~\cite{AMEGO:2019gny,e-ASTROGAM:2016bph} would provide a powerful probe of the physical nature of this intriguing source.

\section{A Model for the Neutrino and Gamma-Ray Emission from NGC 1068}
\label{sec:model}

In this section, we describe our physical model for the neutrino and gamma-ray emission from NGC 1068. This model includes the acceleration and diffusion of high-energy protons, and their subsequent production of gamma rays, electrons, and neutrinos. We also consider the interactions of gamma rays and electrons through the processes of pair production, synchrotron, and inverse Compton scattering.

\subsection{Proton Acceleration and Diffusion}

We take the spectrum of protons in the corona of the AGN to be given by
\begin{equation}
    \left(\dfrac{\de N_p}{\de E_p} \right)_\drm{inj} \propto \left[1 - \exp \left(-\dfrac{E_p}{m_p c^2} \right) \right] \, \left(\frac{E_p}{1 \, \rm GeV}\right)^{-\Gamma_p} \, \exp\left(-\dfrac{E_p}{E^{\rm{max}}_p} \right) \; ,
\end{equation}
%
where $E^{\rm{max}}_p$ is the maximum energy to which protons are accelerated. This can be estimated by finding the energy at which the timescale for acceleration is equal to that of escape. 
Since the transport of protons through the corona is dominated by diffusion, we follow Ref.~\cite{Murase:2019vdl} to make the following estimate for the time required for a proton to escape from the corona~\cite{Dermer:1995ju,Dermer:2014vaa,Kimura:2014jba,Murase:2011cx,Stawarz:2008sp}:
\begin{equation}
    t_\drm{diff} = \dfrac{3\sqrt{3} \, R}{\eta c} \left(\dfrac{e B R}{\sqrt{3} \, E_p} \right)^{2-q} \; ,
    \label{eq:diff}
\end{equation}
%
where $R$ is the radius of the corona, $B$ is the magnetic field strength in the corona, and $q$ is the spectral index in momentum space of the particles accelerated by a stochastic magnetic field. In our analysis, we take $q=5/3$, corresponding to the case of Kolmogorov diffusion. The quantity $\eta$ is related to the power spectrum, $P_k$, of the turbulence via $\eta = [8 \pi \int dk P_k/B^2]^{-1}$, where $P_k \propto k^{-q}$~\cite{Murase:2019vdl}. 
The radius of NGC 1068's corona is estimated to be on the order of $R \sim (3-100) \, R_s$~\cite{Murase:2019vdl}, where $R_s = 2 GM_{\rm BH}/c^2$ is the Schwartzchild radius of the galaxy's supermassive black hole. 

The timescale for particle acceleration is given by 
\begin{equation}
t_\drm{acc} =  \dfrac{R \eta c}{\sqrt{3} \, v^2_a} \left(\dfrac{\sqrt{3} \, E_p}{e B R} \right)^{2-q} \; ,
\label{eq:acc}
\end{equation}
where $v_A = B/\sqrt{4 \pi m_p n_p}$ is the Alfvén velocity. The nucleon density in the corona is \mbox{$n_p = \sqrt{3} \,\tau_T / \left(\sigma_T R \right)$}, where $\tau_T$ and $\sigma_{T}$ are the Thomson optical depth and cross section, respectively. 

Equating the diffusion and acceleration timescales as given in Eqs.~\ref{eq:diff} and~\ref{eq:acc} and solving for energy, we find 
\begin{align}
E^{\rm max}_p &= \bigg(\frac{3 v_a}{\eta c}\bigg)^{1/(2-q)} \, \frac{eBR}{\sqrt{3}} \\
&= \frac{3^{7/4} \, e B^4 R^{5/3} \sigma_T^{3/2}}{8 \pi^{3/2} \eta^3 c^3  m_p^{3/2} \tau_T^{3/2}}  \nonumber \\
&\approx 3 \times 10^5 \, {\rm GeV} \times \bigg(\frac{B}{10 \, {\rm kG}}\bigg)^4 \, \bigg(\frac{R}{25 \, R_s}\bigg)^{5/3} \,  \bigg(\frac{M_{\rm BH}}{2\times 10^7  \, M_{\odot}}\bigg)^{5/3}\,\bigg(\frac{25}{\eta}\bigg)^3 \, \bigg(\frac{0.5}{\tau_T}\bigg)^{3/2} \; , \nonumber
\end{align}
where in the second and third lines we have taken $q=5/3$.

\subsection{Pion Production}

The optical depth for inelastic proton-proton scattering is given by
\begin{equation}
\tau_{pp}(E_p) = n_p k_{pp} \sigma_{pp}(E_p) \; c \; t_\drm{diff}(E_p) \; ,
\end{equation}
where $k_{pp} \approx 0.5$ \cite{Musulmanbekov:2003wy} is the average proton inelasticity of such interactions and $\sigma_{pp}$ is the total inelastic cross section for proton-proton scattering (for a useful parameterization, see Ref.~\cite{Kamae:2006bf}). Since $\sigma_{pp}$ is roughly constant over the energy range of interest and $t_{\rm diff} \propto E_p^{-1/3}$, the optical depth is expected to fall with energy, $\tau_{pp} \propto E_p^{-1/3}$. 

These interactions result in the production of charged and neutral pions, which produce neutrinos and gamma rays through their respective decays. The average number of pions that is produced in such a collision scales as $N_{\pi} \propto E_p^{1/4}$, while the average fraction of the energy that is carried by a given pion scales as $\langle E_{\pi}/E_p \rangle \sim 0.03 \times (E_p/{\rm TeV})^{-1/4}$~\cite{BeckerTjus:2014uyv,Mannheim:1994sv}. These factors cause the neutrinos and gamma rays to have a softer spectral index than their parent protons, $\Gamma_{\nu, \gamma} \approx -(4/3)\Gamma_p +(2/3)$~\cite{BeckerTjus:2014uyv}. Combining this with the energy dependence of the optical depth, we arrive at the following spectra of neutrinos, electrons, and gamma rays:
\begin{align}
\frac{dN_{\nu}}{dE_{\nu}} &\propto E_{\nu}^{-\frac{4}{3}\Gamma_p+\frac{2}{3}} \, e^{-E_{\nu}/E^{\rm max}_ \nu} \, \big(1-e^{-\tau_{pp}(E_p)}\big)\\
\frac{dN_e}{dE_e} &\propto E_{e}^{-\frac{4}{3}\Gamma_p+\frac{2}{3}} \, e^{-E_{e}/E^{\rm max}_{e}} \, \big(1-e^{-\tau_{pp}(E_p)}\big) \nonumber \\
\frac{dN_{\gamma}}{dE_{\gamma}} &\propto E_{\gamma}^{-\frac{4}{3}\Gamma_p+\frac{2}{3}} \, e^{-E_{\nu}/E^{\rm  max}_{\gamma}} \, \big(1-e^{-\tau_{pp}(E_p)}\big) \; , \nonumber
\end{align}
where $E^{\rm max}_{\nu} = E^{\rm max}_{e} \sim 0.025 \,E^{\rm max}_{p}$ and $E^{\rm max}_{\gamma} \sim 0.05 \, E^{\rm max}_{p}$. These spectra are normalized such that
\begin{align}
     \int E_{\gamma} \, \frac{dN_{\gamma}}{dE_{\gamma}}  \, dE_\gamma &= \frac{1}{3}\int E_p \,\frac{dN_p}{dE_p} \, (1-e^{-\tau_{pp}}) \, dE_p \\
     \int E_{e} \, \frac{dN_{e}}{dE_{e}}  \, dE_e &= \frac{1}{6}\int E_p \,\frac{dN_p}{dE_p} \, (1-e^{-\tau_{pp}}) \, dE_p \nonumber \\
     \int E_{\nu} \, \frac{dN_{\nu}}{dE_{\nu}}\, dE_{\nu} &= \frac{1}{2}\int E_p \,\frac{dN_p}{dE_p} \, (1-e^{-\tau_{pp}}) \, dE_p. \nonumber
\end{align}
The prefactors of $1/3$, $1/6$, and $1/2$ come from the facts that two charged pions are produced for every neutral pion in proton-proton collisions, and that three of the four decay products of a charaged pion are neutrinos. Note that in the $\tau_{pp} \ll 1$ limit, $1-e^{-\tau_{pp}} \approx \tau_{pp} \propto E_p^{-1/3}$, and thus the neutrino and gamma-ray spectra take on power-law indices of $\Gamma_{\nu, \gamma} \approx -(4/3)\Gamma_p +(1/3)$.

\subsection{Gamma-Ray Attenuation}

In order to explain the lack of very high-energy gamma-ray emission observed from NGC 1068, energetic photons must be efficiently attenuated through the process of pair production. The optical depth for such interactions is given by
\begin{equation}
    \tau_{\gamma \gamma}(E_{\gamma}) =  R \int^1_{-1} \frac{1-\cos \theta}{2} d(\cos \theta) \int  \frac{dn_{\rm rad}}{d\epsilon_{\rm rad}}(\epsilon_{\rm rad}) \, \sigma_{\gamma \gamma}(E_{\gamma}, \epsilon_{\rm rad},\theta)  \, d\epsilon_{\rm rad} \; ,
\end{equation}
where $dn_{\rm rad}/d\epsilon_{\rm rad}$ is the differential number density of photon targets in the corona and $\sigma_{\gamma \gamma}$ is the cross section for pair production~\cite{Breit:1934zz,Gould:1967zzb,1983Afz....19..323A}:
\begin{align}
\label{eq:pairprod}
\sigma_{\gamma \gamma} = \frac{2 \pi \alpha^2}{E_{\rm CM}^2}  \bigg[2\beta (\beta^2-2)+(3-\beta^4) \ln\bigg(\frac{1+\beta}{1-\beta}\bigg)\bigg] \; ,
\end{align}
where $\beta = [1-(2m_e/E_{\rm CM})^2]^{1/2}$ and $E_{\rm CM} = [2E_{\gamma} \epsilon_{\rm rad} \, (1-\cos \theta)]^{1/2}$ is the energy of the collision in the center-of-momentum frame. For the radiation field, we adopt a two-component model~\cite{Fang:2022trf} consisting of a $T=4.2 \, {\rm eV}$ black-body spectrum and a $E^{-2}$ power-law with an exponential cutoff above 130 keV. This later component is normalized such that the X-ray luminosity integrated between 2 and 10 ${\rm keV}$ is equal to the measured value from NGC 1068, $L_X= 7 \times 10^{43} \, {\rm erg/s}$~\cite{Marinucci:2015fqo,Bauer:2014rla}.

\begin{figure}[t]
\centering
\includegraphics[width=0.60\textwidth]{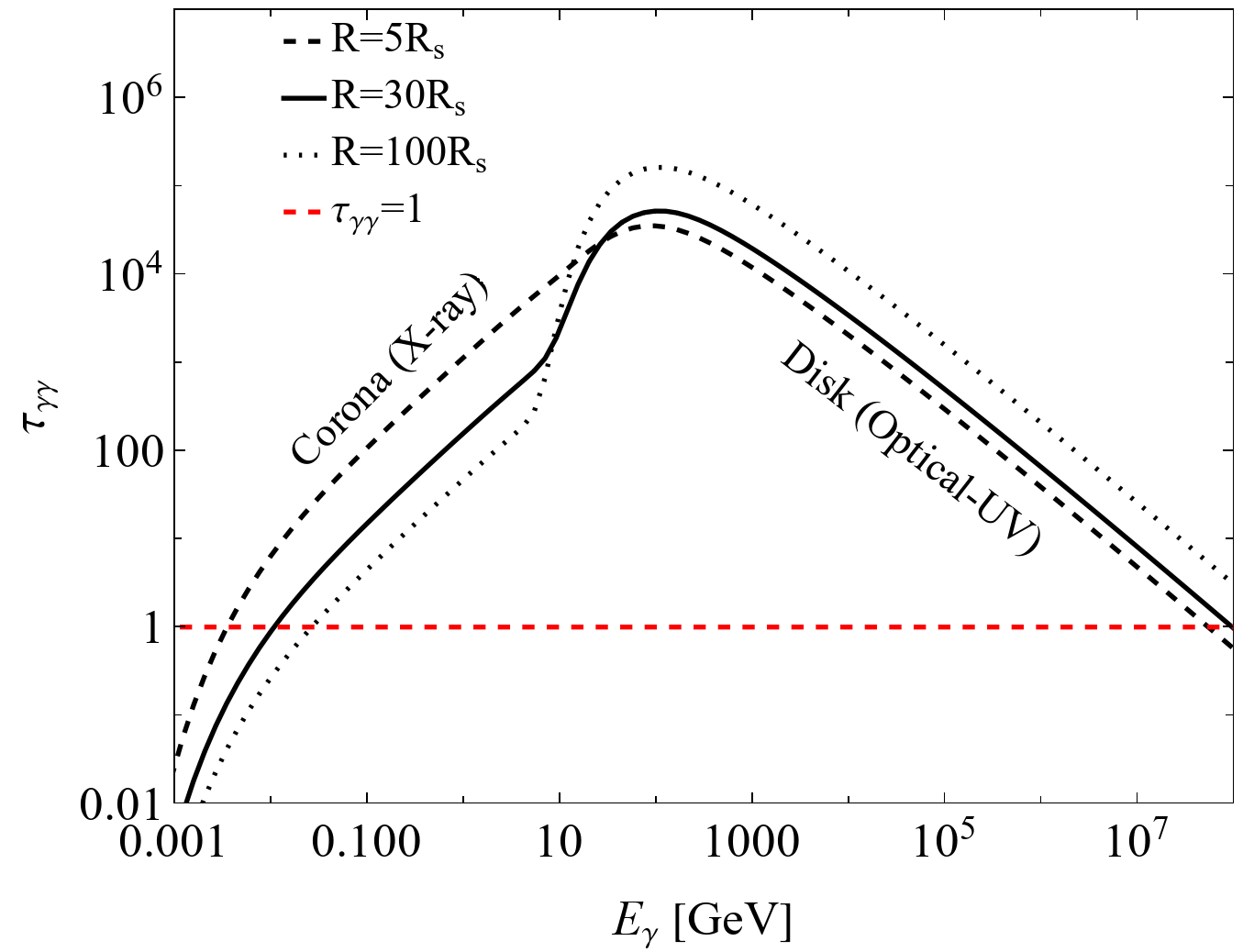}
\caption{The optical depth of the NGC 1068's corona to pair production for three values of the radius, $R$. At high energies, the corona is optically thick, but is transparent to photons with energies below $E_{\gamma} \sim \mathcal{O}(10 \, {\rm MeV})$.}
\label{fig:opticaldepth}
\end{figure}

In Fig.~\ref{fig:opticaldepth}, we show the optical depth of NGC 1068's corona to pair production as a function of the gamma-ray energy, for three choices of $R$. At high energies, the corona is very optically thick, but is transparent to photons with energies below $E_{\gamma} \sim \mathcal{O}(10 \, {\rm MeV})$.

\subsection{Synchrotron and Inverse Compton Emission}

High-energy electrons interact with the magnetic and radiation fields in the corona to produce synchrotron and inverse Compton emission, respectively. The emissivity of these signals (which has units of energy per volume per time) are found by convolving the electron number density with the power radiated via synchrotron or inverse Compton per electron:
\begin{equation}
    j_{\rm IC/Syn}(E_\gamma) = \int \mathcal{P}_{\rm IC/Syn} \left(E_\gamma, E_e \right) \dfrac{\de n_e}{\de E_e} \left(E_e \right)  \, \de E_e \; .
\end{equation}
The differential number density of electrons, $dn_e/dE_e$, can be found by solving the transport equation. For the energies of interest, the diffusion timescale is always much longer than the timescale of the energy losses, thus the 
electron number density reduces to
%
%
%
%
%
\begin{equation}
    \dfrac{\de n_e}{\de E_e} \left(E_e \right) = \dfrac{1}{b_{\rm{IC}}+ b_{\rm{Syn}}} \int_{E_e}^{\infty} \de \Tilde{E_e} \, Q_e( \Tilde{E_e}) \; ,
    \label{dnde}
\end{equation}
where $b_{\rm IC}$ and $b_{\rm syn}$ and the energy loss rates due to the process of inverse Compton scattering and synchrotron. The injection term, $Q(\Tilde{E_e})$, denotes the rate per energy and volume at which electrons are injected into the corona:
\begin{align}
Q(E_e) &= 2c  \int \frac{1-\cos \theta}{2} d(\cos \theta) \int \int \frac{dn_\gamma}{dE_\gamma}(E_{\gamma}) \,\frac{dn_{\rm rad}}{d\epsilon_{\rm rad}}(\epsilon_{\rm rad}) \, \frac{d\sigma_{\gamma \gamma}}{dE_e}(E_{\gamma}, \epsilon_{\rm rad}, \theta, E_e) \, dE_{\gamma} d\epsilon_{\rm rad} \; ,
\end{align}
where $d\sigma_{\gamma \gamma}/dE_e$ is the 
differential pair production cross section, parametrizing the probability of generating an electron of a given energy~\cite{1983Afz....19..323A}. 
The rates for synchrotron and inverse Compton  energy losses are given by
\begin{align}
    b_{\rm syn}(E_e, B) &= \dfrac{2\sigma_{\rm T}   B^2 E^2_e}{3 \mu_0 m_e^4 c^7} \\
    b_{\rm IC} (E_e) &=\int  \frac{\mathcal{P}_{\rm IC}(E_e,E_\gamma)}{E_\gamma} dE_\gamma \; , \nonumber
\end{align}
where $\mu_0$ is the vacuum permeability. The differential power in inverse Compton scattering, $\mathcal{P}_{\rm IC}$, is given by
\begin{equation}
    P_{\rm IC} (E_e,E_\gamma) = c \int d \epsilon_{\rm rad}\,\frac{dn_{\rm rad}}{d\epsilon_{\rm rad}}(\epsilon_{\rm rad}) \, E_{\gamma} \, (E_\gamma - \epsilon_{\rm rad}) \, \frac{d\sigma_{\rm IC}}{d E_\gamma}\left(E_\gamma,\epsilon_{\rm rad}, E_e \right) \; ,
\end{equation}
where $d\sigma_{\rm IC}/dE_\gamma$ is the differential IC cross section, parametrizing the probability of generating a secondary photons of a given energy~\cite{1981Ap&SS..79..321A}. The analogous quantity for synchrotron is
\begin{equation}
    P_{\rm syn} \left(E_e, \nu \right) = \dfrac{\sqrt{3} \, e^3 B}{4 \pi \epsilon_0 m_e c} G (\nu/\nu_c) \; ,
\end{equation}
where the critical frequency is $\nu_c =  3 e B E_e^2 /4 \pi m^2_e c^2$ and the $G$-function can be approximated by \cite{Fornengo:2011iq} \footnote{The interested reader can check that this is a good approximation of $G\left(x \right) = 2 \pi \int_0^{2\pi} {\rm d \theta } \,{\rm sin}{\theta} \, F \left(\dfrac{x}{{\rm sin}\theta} \right)$ where $F \left( x \right) = x \int_{x}^\infty {\rm d y} K_{5/3} \left(y \right)  $ and $ K_{5/3}$ is a modified Bessel function of second kind and order $5/3$.} 
\begin{equation}
    G(x) = a x^d \, \exp \left(- \sqrt{x/b} - x/c \right) \; ,
\end{equation}
where $a= 1.609$, $b=1.959$, $c=1.131$, and $d=0.3384$.

From the synchrotron and inverse Compton emissivities, we can calculate the gamma-ray flux per solid angle, as observed at Earth:
\begin{align}
\frac{dN_{\gamma}}{dE_{\gamma}}(E_{\gamma}) &=\frac{1}{E_{\gamma}^2} \int_{\rm los}\frac{j_{\rm IC}(E_{\gamma})+j_{\rm syn}(E_{\gamma})}{4\pi} \, ds \; ,
\end{align}
where the integral is performed over the line-of-sight. Integrating over the solid angle, we obtain the following differential flux (per unit of energy, time, and area):
\begin{align}
\frac{dN_{\gamma}}{dE_{\gamma}}(E_{\gamma}) =  \frac{[j_{\rm IC}(E_{\gamma})+j_{\rm syn}(E_{\gamma})] R^3}{3 d_L^2 E_{\gamma}^2} \; ,
\end{align}
where $d=12.7 \, {\rm Mpc}$ is the distance to NGC 1068~\cite{ricci2017bat}.

\section{\Fermi-LAT Data Analysis}
\label{sec:analysis}

\begin{figure}[t]
\centering
\includegraphics[width=0.6\textwidth]{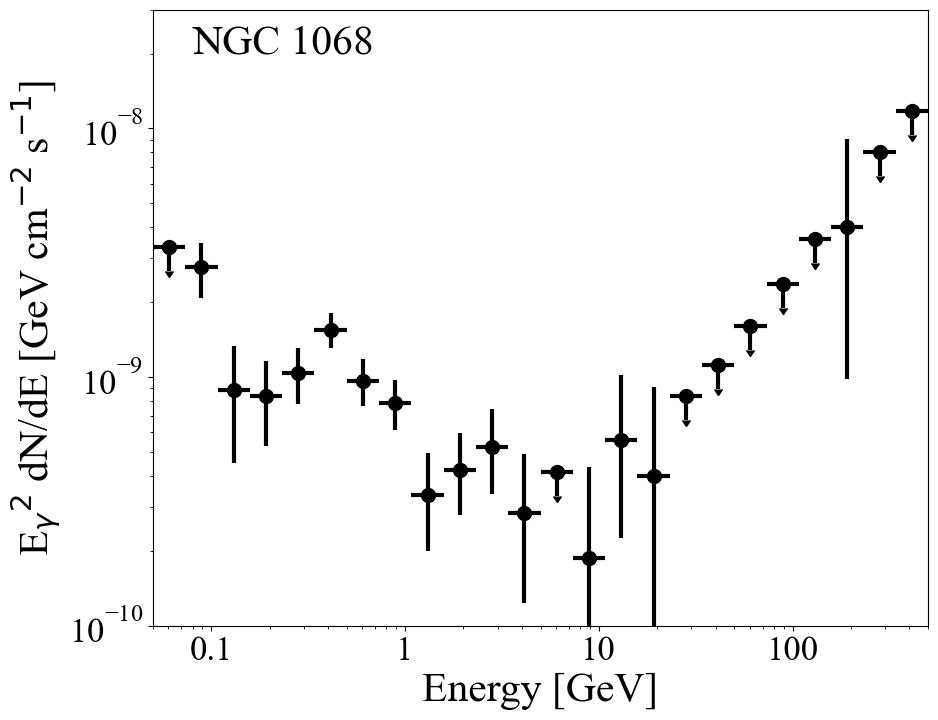}
\caption{The gamma-ray emission from NGC 1068, as measured by the {\sl Fermi Gamma-Ray Space Telescope}. The gamma-ray flux from this source is soft but bright at energies below $\sim 1 \, {\rm GeV}$.}
\label{fig:FermiNGC1068}
\end{figure}

We analyzed nearly 15 years of \Fermi-LAT data\footnote{Mission Elapsed Time (MET): 239557417 -- 707477412.} from a 20$^\circ$$\times$20$^\circ$ region of interest surrounding NGC 1068, as well as surrounding the active galaxies NGC 4151 and NGC 3079. To this end, we utilized the standard {\tt fermipy} package which, in turn, applies standard cuts on the \Fermi-LAT data as executed in the {\tt Fermitools}~\cite{2017ICRC...35..824W}. More specifically, we include SOURCE class events observed in both the front and the back of the detector(evlass=128, evtype=3). We bin the data into 0.1$^\circ$ angular bins and 24 equally spaced logarithmic energy bins spanning from 50~MeV to 500~GeV. We fit the data utilizing the standard {\tt fermipy} algorithm, and additionally we use the algorithm {\tt findsources} to identify and model any fluctuation with $\sqrt{TS} > 3.0$ that is found within a minimum separation of 0.25$^\circ$ from the active galaxy in question. We fit the spectral energy distribution (SED) of each galaxy while allowing the normalization of all sources within 7$^\circ$ to float under the restriction of {\tt cov\_scale=5.0}.

In Fig.~\ref{fig:FermiNGC1068}, we plot the gamma-ray spectrum that we obtain in our analysis from the direction of NGC 1068. We detect the NGC 1068 source with a test statistic (TS) of 255 and a power-law spectral index of $-2.37 \pm 0.07$, although there is also some evidence of a peak-like feature near $E_{\gamma} \sim 0.5 \, {\rm GeV}$. We note that these values are consistent with results reported by the Fermi-LAT collaboration for the source 4FGL J0242.6-0000, which is associated with NGC 1068. In the 4FGL-DR2 catalog, the Fermi-LAT collaboration reports a slightly higher significance (${\rm TS}=337$) and a similar spectral index of $-2.34 \pm 0.05$. 


To assess the robustness of the gamma-ray spectrum we measure from NGC 1068, we have carried out our analysis while making a series of alternative analysis choices. In particular, we have restricted our analysis to include only gamma rays that belong to the upper quartile of event reconstruction (evtype=32). This allows us to analyze only the photons that have the best reconstructed angular distribution, and decreases the covariance between sources and diffuse emission at low energies (at the cost of throwing out 75\% of the total data). Second, we have also reduced the region of interest to 15$^\circ \times 15^{\circ}$, which decreases the effect of distant diffuse emission modeling errors at the cost of allowing bright distant sources to affect the low-energy analysis. Third, we have considered an alternative energy range of 40~MeV -- 400~GeV, which redistributes any features that might exist at the edges of our energy bins. Fourth, we have fixed the spectrum of distant point sources to their best fit values while fitting the spectrum from NGC 1068, decreasing the number of free-parameters in the NGC 1068 spectral fit. Fifth, we have alternatively allowed the spectral parameters of the sources within $7^{\circ}$ of NGC 1068 to be completely free ({\tt cov\_scale=0.0}) during the fitting of the NGC 1068 spectrum. This allows us to more accurately fit spectral features in nearby sources that may contaminate the spectral measurement of NGC 1068. In each of these cases, we find that our results are not substantially different from those obtained using our default analysis choices. The only exception is in the lowest energy bin (50.0 -- 73.4~MeV), where fixing other sources during SED fitting changes the upper limit (shown in our default analysis) into a weak $\sim$2$\sigma$ detection of low-energy emission. Thus we consider the results of this energy bin to not be robust and to serve only as a lower limit. 

We repeated the same analysis on the sources NGC 4151 and NGC 3079, which have not yet been detected in the Fermi-LAT catalogs, although we note that NGC 4151 has recently been detected at the level of 5.5$\sigma$ in Ref.~\cite{Peretti:2023crf} with a spectral index of 2.39 $\pm$0.18. In the case of NGC 4151, we also find evidence for gamma-ray emission from this source, with a TS value of 39.6 ($\sim$6.3$\sigma$) and a best-fit flux of $3.9 \times 10^{-10} \, {\rm GeV} \, {\rm cm}^{-2}\,{\rm s}^{-1}$ at 0.1 GeV, and a spectral index of 2.08 $\pm$ 0.21. Our detection is consistent with the result of Ref.~\cite{Peretti:2023crf}. We note that the slightly harder spectral index in our analysis is driven by the inclusion of strong upper limits on the NGC 4151 gamma-ray flux between 50--100~MeV (Ref.~\cite{Peretti:2023crf} use a minimum energy cutoff of 100~MeV). In the case of NGC 3079, we find very weak evidence for a tentative source, with a TS of 13.4 (this lies below the TS~=~25 threshold used by the Fermi collaboration for source detection, but potentially represents a $\sim$3$\sigma$ detection of gamma-ray emission), a best-fit flux of $2.8 \times 10^{-10} \, {\rm GeV} \, {\rm cm}^{-2}\,{\rm s}^{-1}$ at 0.1 GeV, and a spectral index of 2.19 $\pm$ 0.36.

\section{Results}

Our model for the neutrino and gamma-ray emission from NGC 1068, as described in Sec.~\ref{sec:model}, has the following free parameters: 1) the normalization and spectral index of the accelerated protons, 2) the radius of the corona, and 3) the magnetic field strength in the corona. In addition, we take the Thomson optical depth to be $\tau_T = 0.5$, and the diffusion index to be $q=5/3$ (corresponding to Kolmagorov diffusion). We also relate the value of $\eta$  to that of the magnetic field strength according to $\eta = 10 \times (B/6 \,{\rm kG})^2$~\cite{Murase:2019vdl}.

\begin{figure}[t]
\centering
\includegraphics[width=0.345\textwidth]{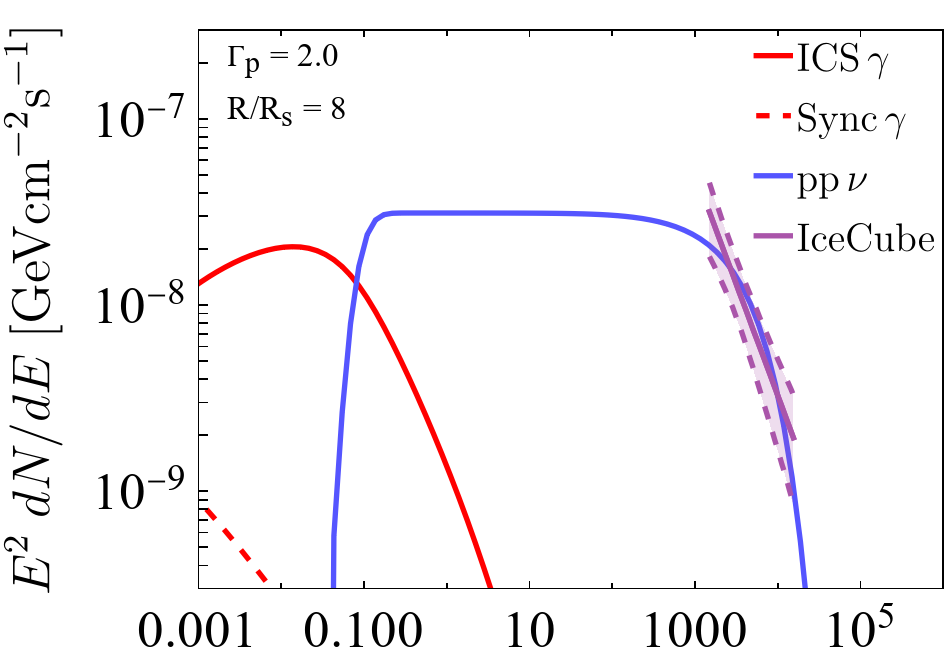}
\includegraphics[width=0.32\textwidth]{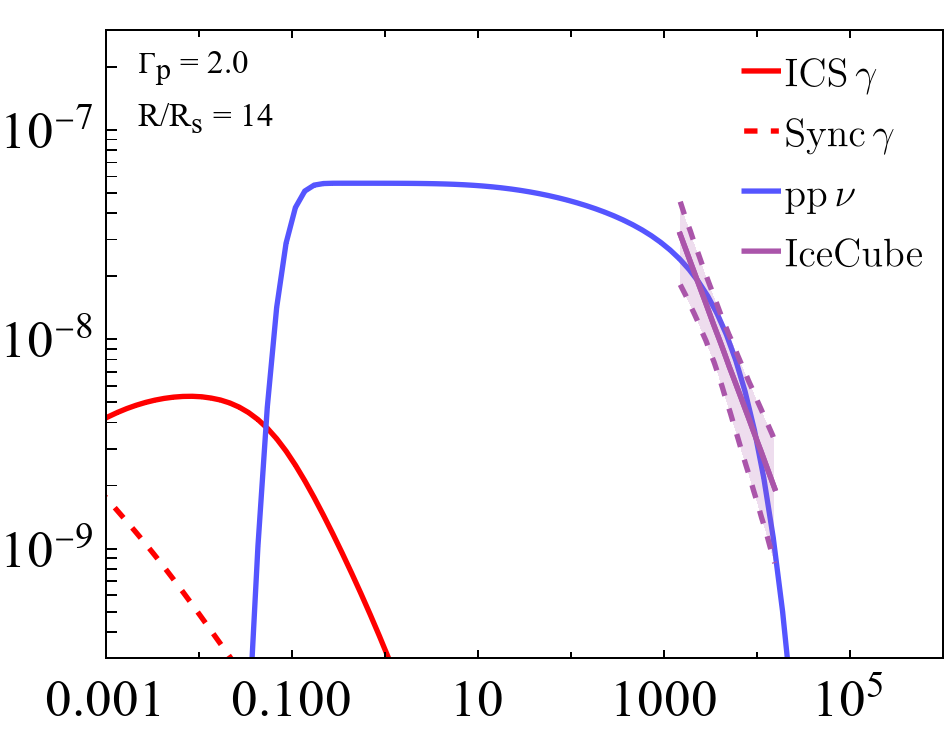}
\includegraphics[width=0.32\textwidth]{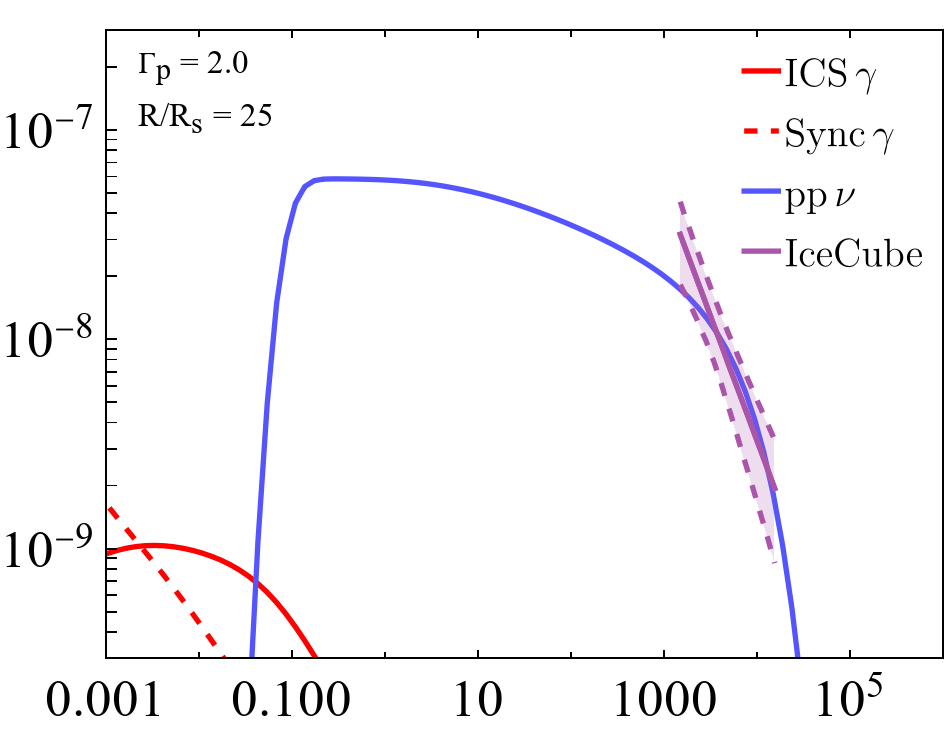}\\
\includegraphics[width=0.345\textwidth]{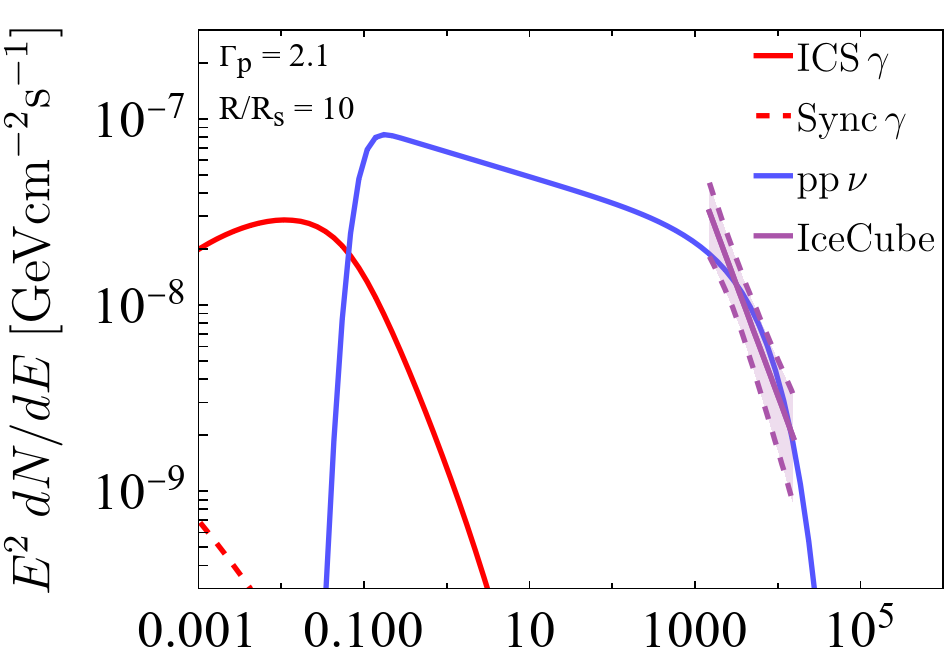}
\includegraphics[width=0.32\textwidth]{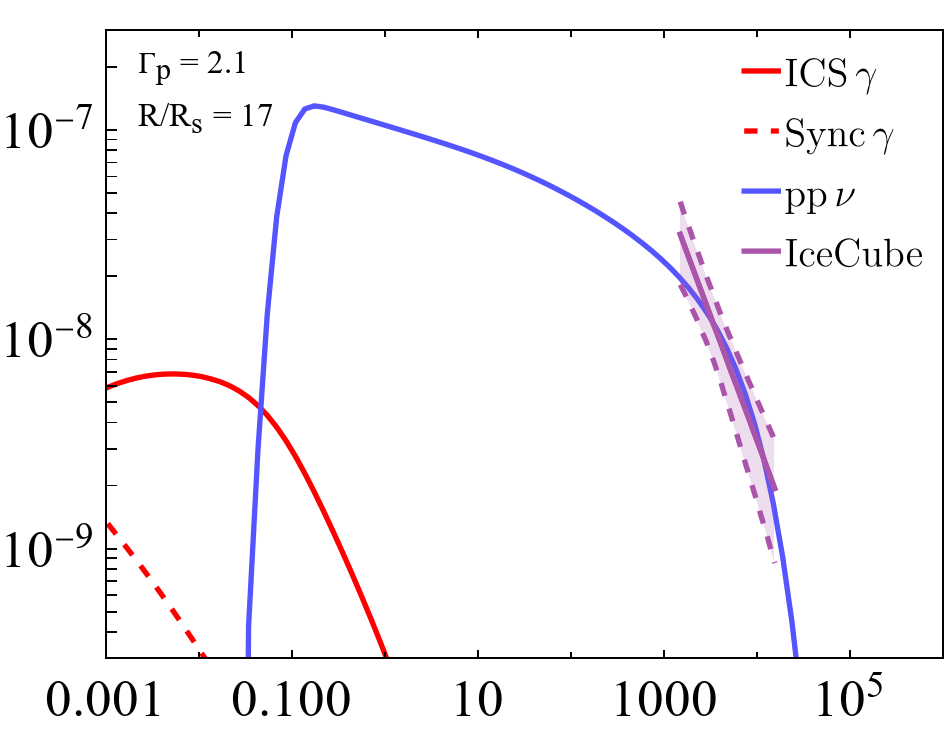}
\includegraphics[width=0.32\textwidth]{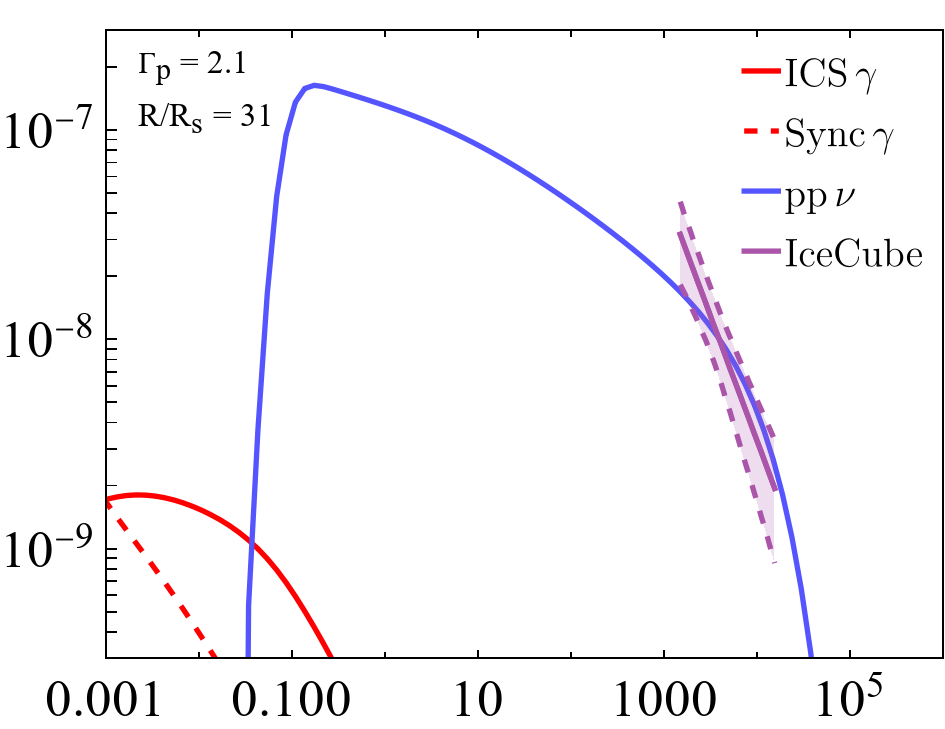}\\
\includegraphics[width=0.345\textwidth]{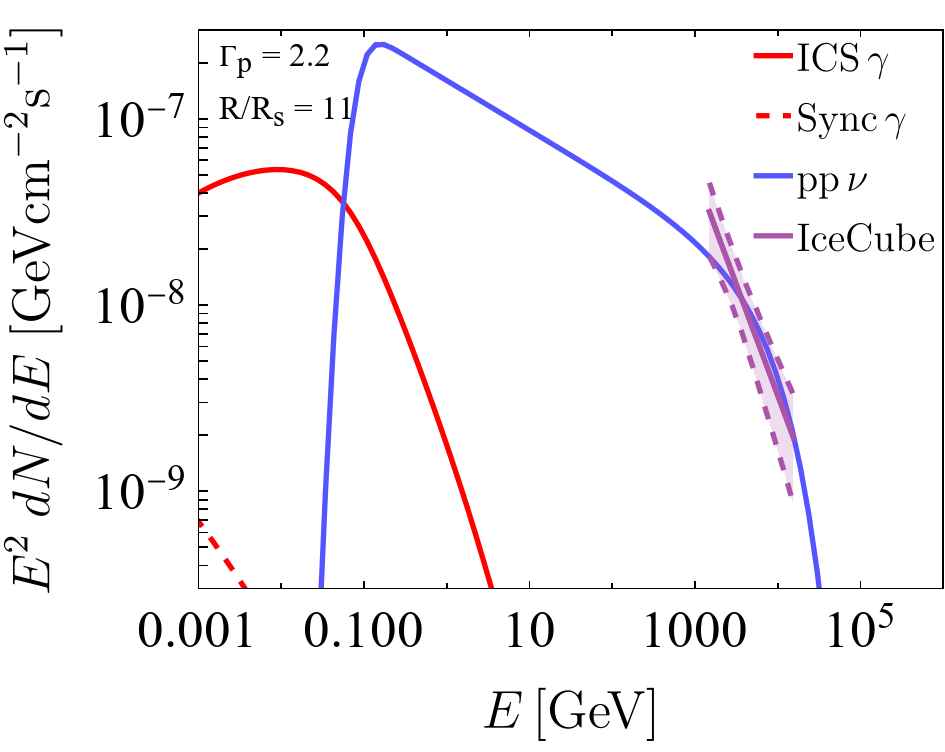}
\includegraphics[width=0.32\textwidth]{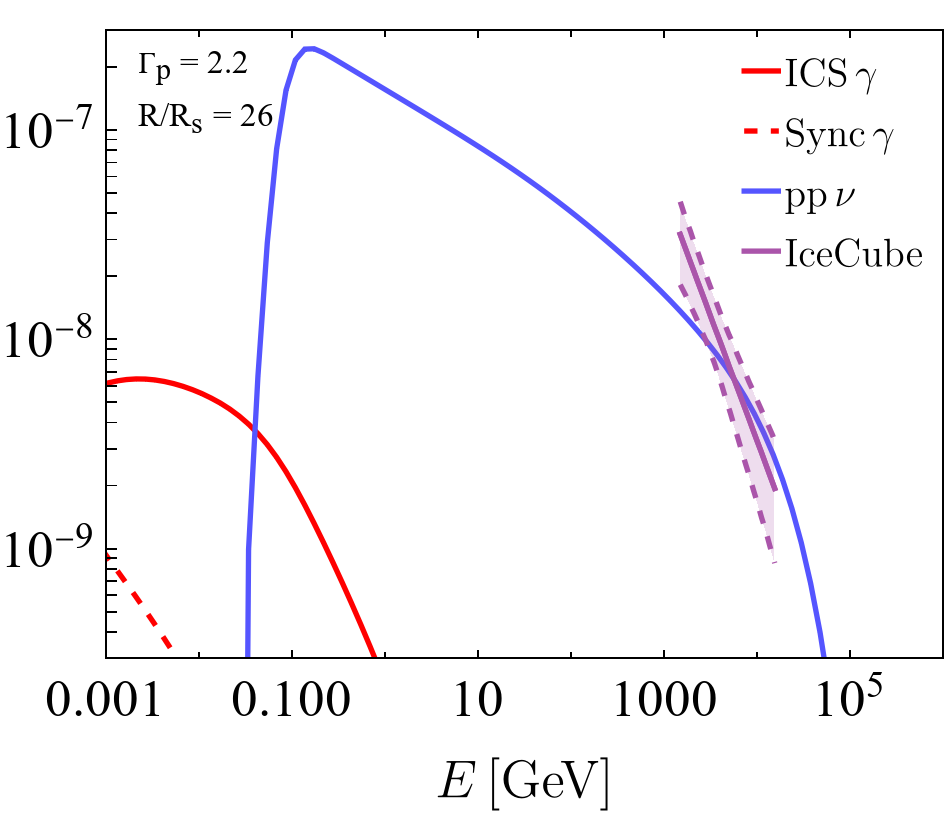}
\includegraphics[width=0.32\textwidth]{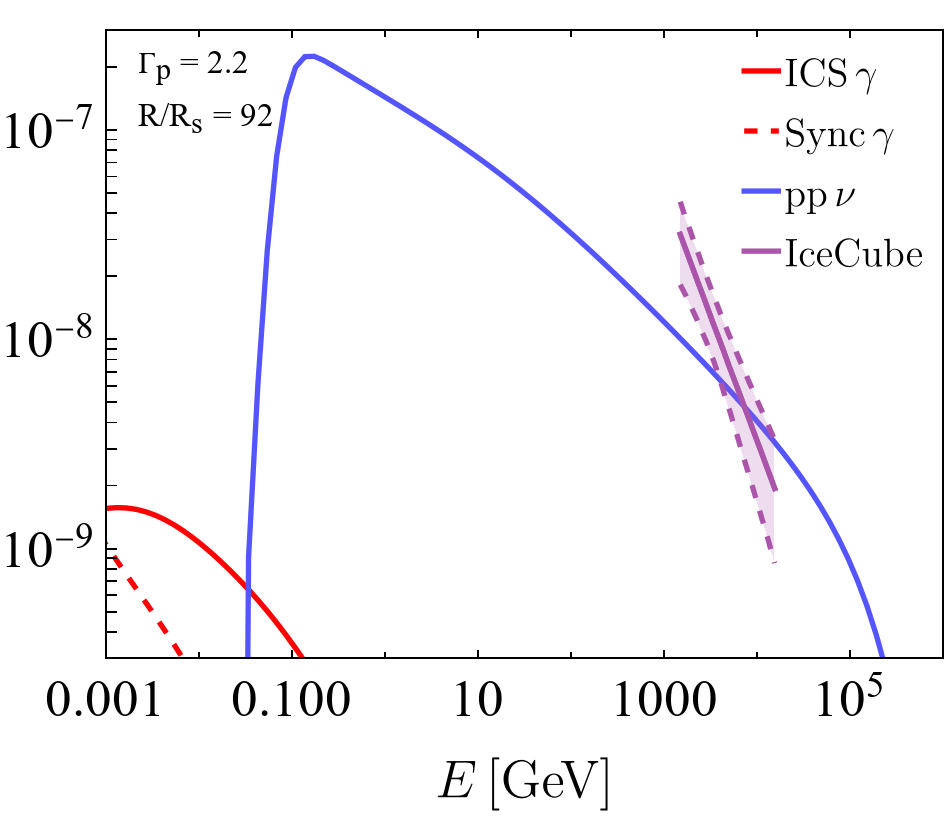}
\caption{The neutrino and gamma-ray emission from NGC 1068 as predicted by our model for various values of the injected proton spectral index,  as well as the magnetic field and radius of the corona. In each frame, we have selected values of $R$ and the overall normalization that provide the best possible fit to the measured neutrino flux. In each case shown, the total power in accelerated protons is comparable to the intrinsic X-ray luminosity of this source, $L_p \approx (0.2-0.8) \, L_X$. To not exceed the sub-GeV gamma-ray flux measured by \Fermi-LAT (see Fig.~\ref{fig:FermiNGC1068}), the magnetic field within the corona of NGC 1068 must be at least $B \gsim 6 {\rm kG}$. From left to right, the magnetic field in each column is taken to be: 3 kG, 6 kG, and 10 kG.}
\label{fig:results}
\end{figure}

The results of our model for NGC 1068 are shown in Fig.~\ref{fig:results} for various choices of the input parameters $\Gamma_p$, $B$, and $R$. In each frame, we have selected values of $R$ and the overall normalization that provide the best possible fit to the neutrino flux reported by the IceCube Collaboration. In all nine cases shown, the total power in accelerated protons is comparable to the intrinsic X-ray luminosity of this source, $L_p \approx (0.2-0.8) \, L_X$.

In each of the three left frames of Fig.~\ref{fig:results} (with $B=3 \, {\rm kG}$), the predicted flux of the gamma rays from inverse Compton scattering are significantly higher than those measured by \Fermi-LAT (see Fig.~\ref{fig:FermiNGC1068}), by a factor of $\sim 5$ at $E_{\gamma}=0.1 \, {\rm GeV}$. In the middle frames (with $B=6 \, {\rm kG}$), the predicted gamma-ray emission approximately matches the one measured by \Fermi-LAT. As a result, we conclude that the magnetic field within the corona of NGC 1068 must be at least $B \gsim 6 {\rm kG}$. Larger values of $B$ are possible, but would require an additional mechanism for the gamma-ray production.

\section{Other Nearby Active Galaxies}

\begin{figure}[t]
\centering

\includegraphics[width=0.49\textwidth]{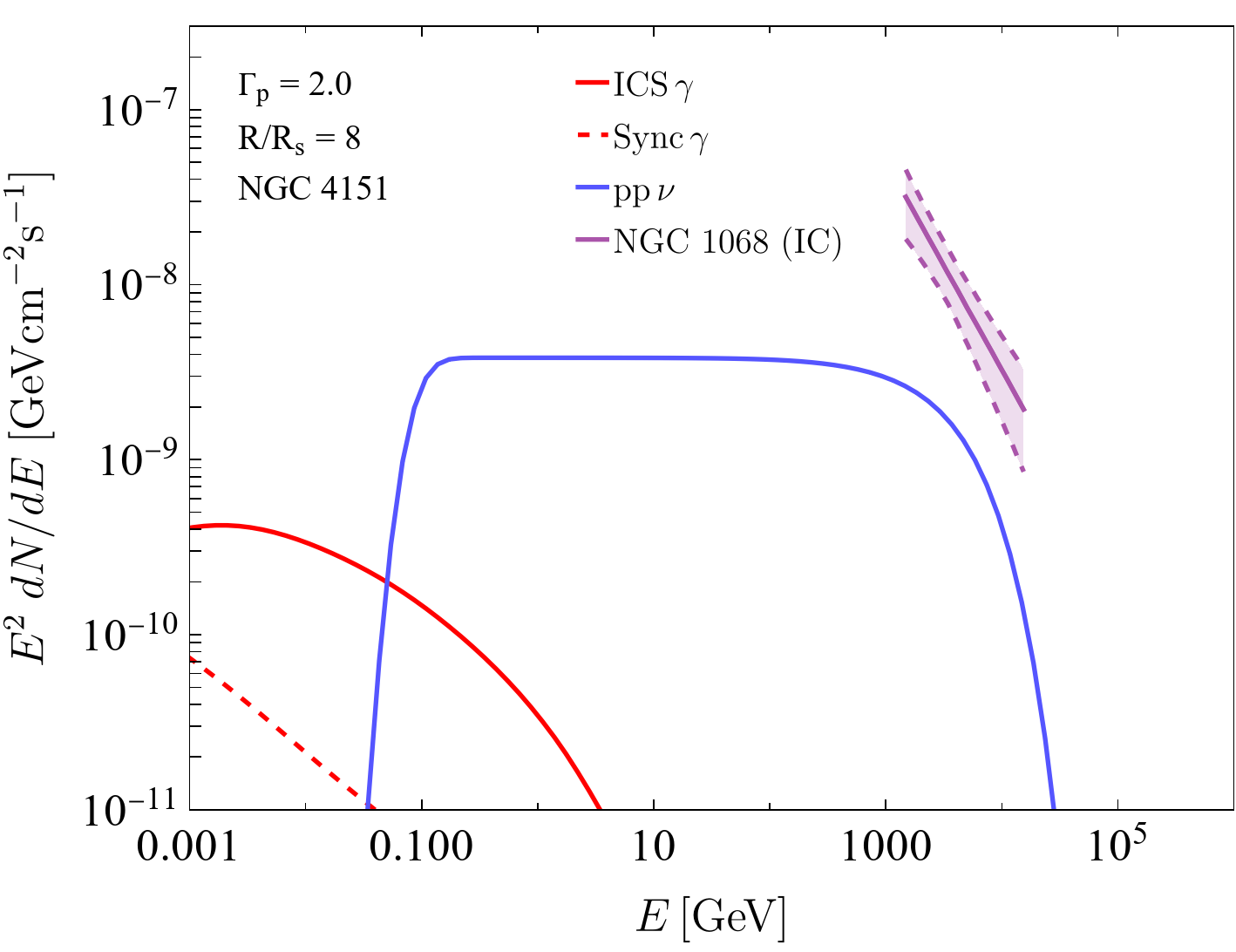}
\includegraphics[width=0.49\textwidth]{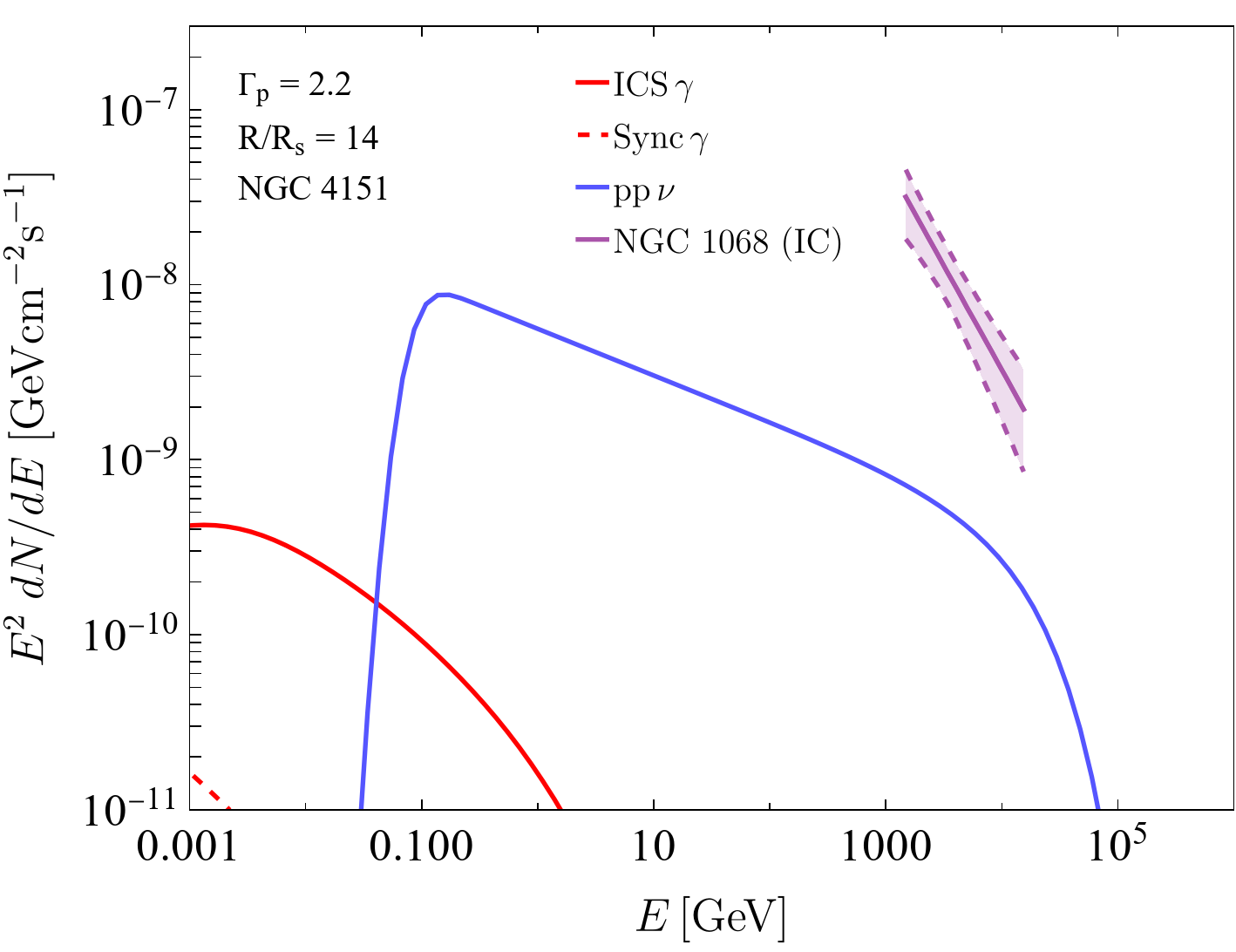}
\includegraphics[width=0.49\textwidth]{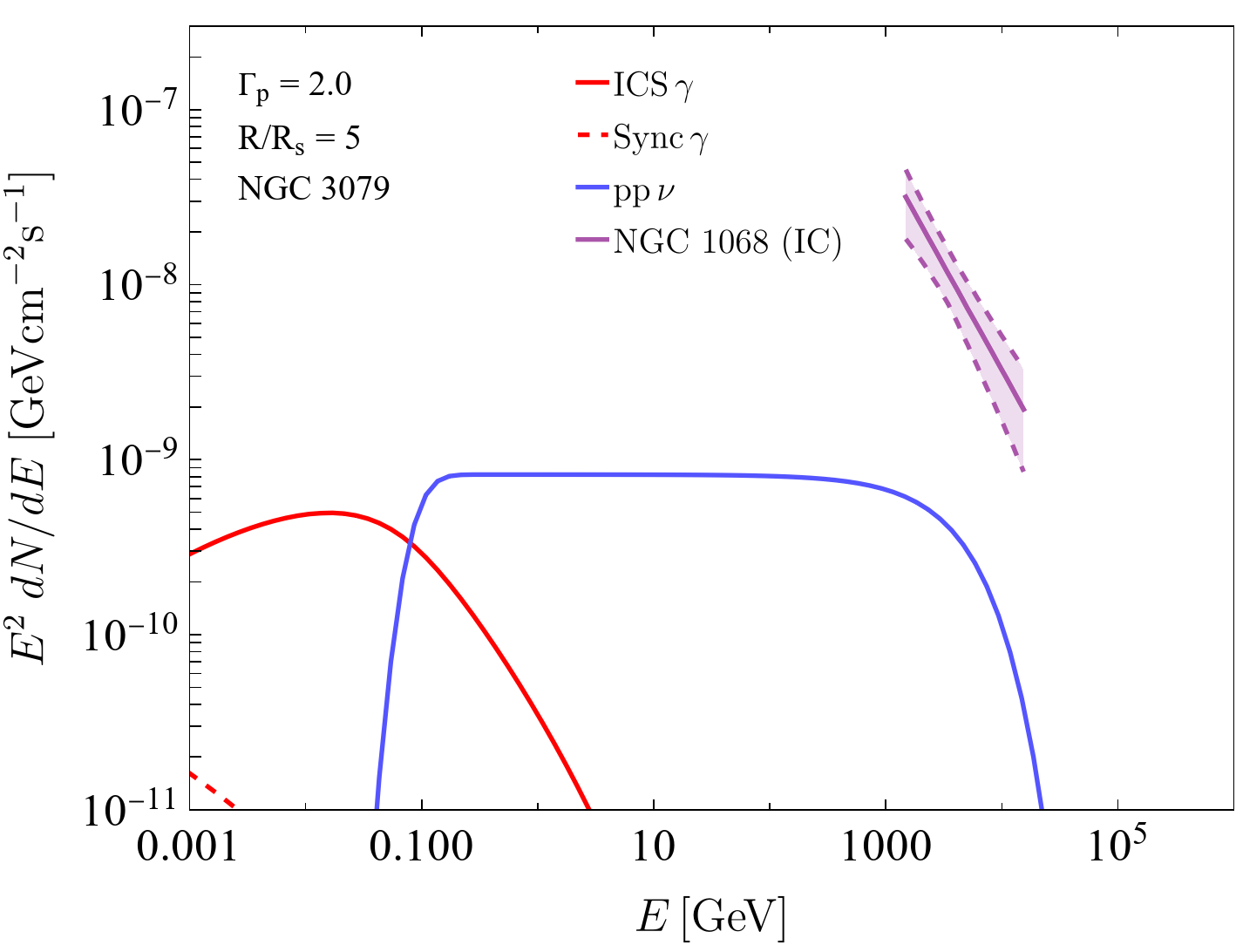}
\includegraphics[width=0.49\textwidth]{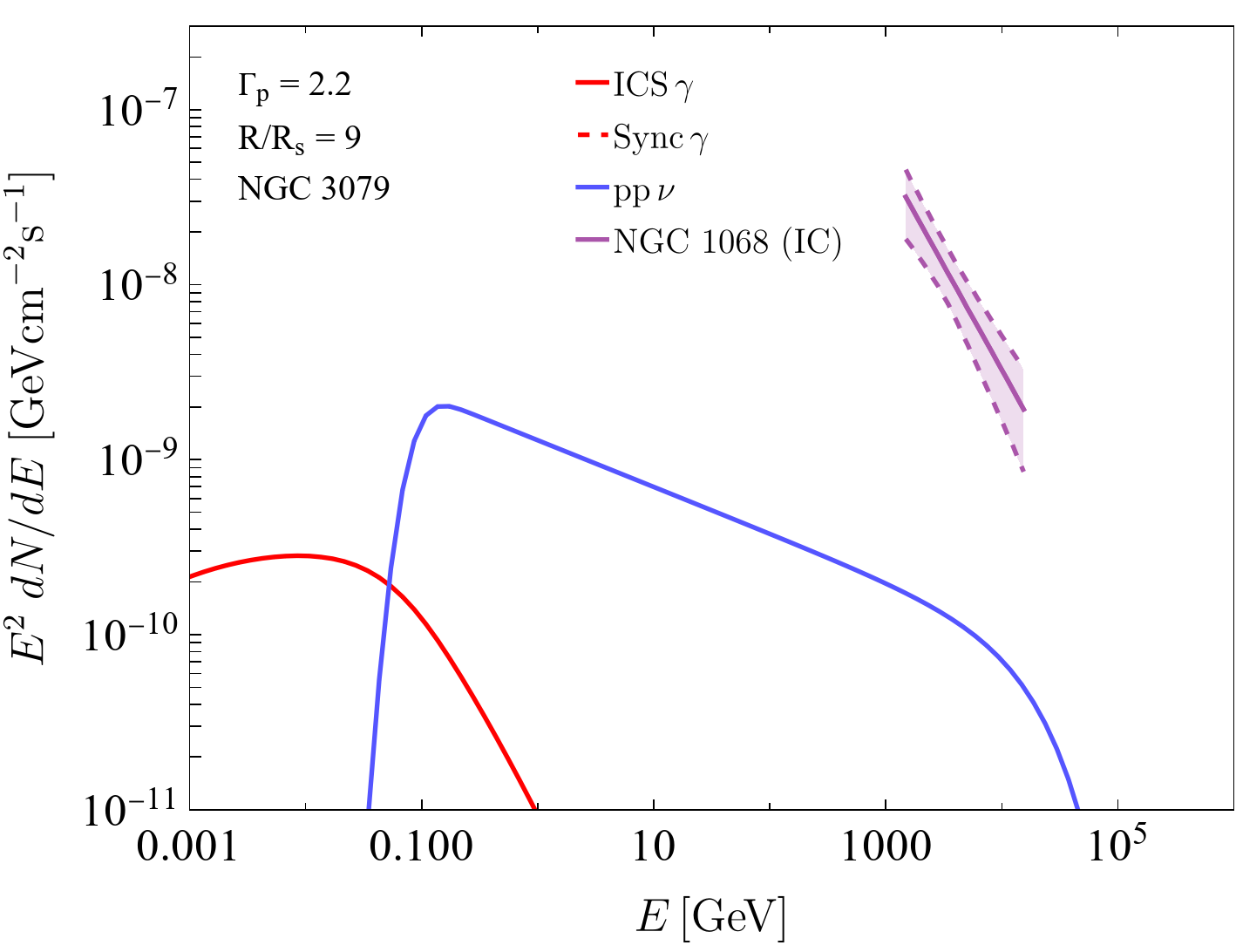}
\caption{The neutrino and gamma-ray emission from NGC 4151 (top) and NGC 3079 (bottom) as predicted by our model for two values of the injected proton index and coronal radius, and for a magnetic field of $B=3 \, {\rm kG}$. These parameter values were chosen to obtain the same value of $E^{\rm max}_p$ as in the top-center and bottom-center frames of Fig.~\ref{fig:results}. For comparison, we show the neutrino emission from NGC 1068, as measured by IceCube.}
\label{fig:other1}
\end{figure}

Motivated by IceCube's detection of high-energy neutrinos from NGC 1068, we have considered the possible neutrino emission from the active galaxies NGC 4151 and NGC 3079 (the two closest X-ray bright, northern hemisphere Seyfert galaxies~\cite{Neronov:2023aks}). In modeling these sources, we have set $L_p/L_X=1$ and adopted the best-fit values of $E^{\rm max}_p$ based on our analysis of NGC 1068, as appearing in the top-center and bottom-center frames of Fig.~\ref{fig:results}. For each galaxy, we use the measured values of the distance, intrinsic X-ray luminosity, and supermassive black hole mass. In particular, for NGC 4151 we adopt $d \approx  12.7\, \rm Mpc$~\cite{ricci2017bat}, $L_X = 2\times 10^{42}\,\rm erg / s$~\cite{ricci2017bat} (see also Refs.~\cite{ackermann2012search,panessa2006x,perola1994x}),  and $M_{\rm BH} = 3\times 10^{7}$~\cite{bentz2015agn,yuan2020cepheid}, while for NGC 3079 we take $d \approx 19.1\, \rm Mpc$~\cite{ricci2017bat}, $L_X = 1\times 10^{42}\,\rm erg / s$~\cite{ricci2017bat,masini2018obscured}, and $M_{\rm BH} = 2\times 10^{6}$~\cite{yuan2020cepheid}. For these parameters and $B=3 \, {\rm kG}$, it follows that $R$ must be less than $\sim 15 \, R_s$ for NGC 4151 and less than $\sim 10 \, R_s$ for NGC 3079.

The neutrino and gamma-ray emission predicted from NGC 4151 and NGC 3079 are shown in Fig.~\ref{fig:other1} for two values of $\Gamma_p$. For comparison, we show the neutrino emission from NGC 1068, as measured by IceCube. For magnetic fields of $B \gsim 1 \, {\rm kG}$, we find that our model can accommodate the gamma-ray flux observed from these sources by \Fermi-LAT, $\sim (3-4) \times 10^{-10} \, {\rm GeV} \, {\rm cm}^{-2}\,{\rm s}^{-1}$ at 0.1 GeV( see Sec.~\ref{sec:analysis}). Future MeV-scale observations of these AGN, such as by AMEGO-X~\cite{AMEGO:2019gny} or e-ASTROGAM~\cite{e-ASTROGAM:2016bph}, could be sensitive to this predicted gamma-ray flux.



\section{Summary and Discussion}

IceCube's recent detection of $\sim 1-10 \,{\rm TeV}$ neutrinos from NGC 1068, combined with the lack of TeV-scale gamma-ray emission from this source, indicates that these neutrinos must be generated in a region that is opaque to very high-energy photons, such as within the dense corona that surrounds NGC 1068's supermassive black hole.

\begin{figure}[t]
\centering
\includegraphics[width=0.6\textwidth]{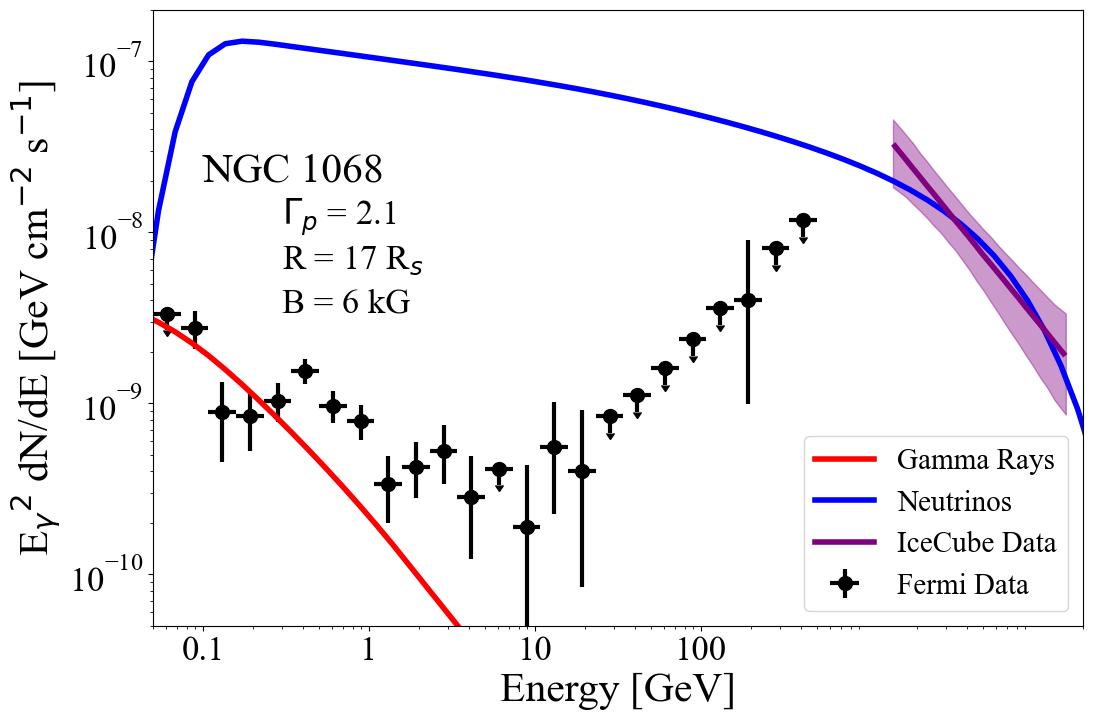}
\caption{The neutrino and gamma-ray emission from NGC 1068 as predicted by our model for a selected value of the injected proton index, $\Gamma_p$, the radius of the AGN's corona, $R$, and magnetic field of $B=6 \, {\rm kG}$. We have chosen the overall normalization to accommodate the neutrino flux measured by IceCube, which requires the total power in accelerated protons to be comparable to the intrinsic X-ray luminosity of this source, $L_p \approx 0.6 \, L_X$. We also show the spectrum of gamma rays observed from this source, which for $B\sim 6 \, {\rm kG}$ can be generated in our model through a combination of synchrotron and inverse Compton scattering.}
\label{fig:summary}
\end{figure}

In this paper, we have presented a physical model for the transport and interactions of high-energy protons, photons, and electrons in the corona of NGC 1068, including the processes of diffusion, pair production, pion production, synchrotron, and inverse Compton scattering. The high densities of X-rays within NGC 1068's corona leads to the efficient absorption of gamma rays, explaining why very high-energy gamma rays have not been detected from this source. When our model's parameters are chosen to accommodate the spectrum of neutrinos reported by IceCube, a significant gamma-ray flux is predicted at MeV-GeV energies, arising from a combination of synchrotron and inverse Compton scattering.

We have also performed a new analysis of the data collected by the \Fermi-LAT telescope from the direction of NGC 1068, finding that this source is very bright at sub-GeV energies and exhibits a soft spectral shape, $dN_{\gamma}/dE_{\gamma}\propto E^{-2.4}_{\gamma}$. In order for this source to not exceed the gamma-ray flux observed by \Fermi-LAT at sub-GeV energies, the vast majority of the energy in high-energy electrons must be lost to synchrotron, requiring the presence of very large magnetic fields, $B \gsim 6 \, {\rm kG}$.

In Fig.~\ref{fig:summary}, we show the neutrino and gamma-ray spectrum from NGC 1068, as predicted by our model for one choice of input parameters. To accommodate the neutrino spectrum reported by IceCube, we require protons to be injected with a spectral index of $\Gamma_p \sim 2.1$ and for the radius of the AGN corona to be $R \sim  17 \, R_s$, where $R_s$ is the Schwartzschild radius of the supermassive black hole. For a magnetic field strength of $B \sim 6 \, {\rm kG}$, the sub-GeV gamma-ray emission observed by \Fermi-LAT can also be explained by this model. Finally, to normalize the neutrino flux to that observed from NGC 1068 requires that this source accelerates protons with a total power that is comparable to its intrinsic X-ray luminosity, $L_p \sim 0.6 \, L_X$.

We have also applied our model to two other X-ray bright and nearby AGN, NGC 4151 and NGC 3079. Finding that the neutrino emission from these sources could be potentially detectable by future high-energy neutrino telescopes. We also find that these sources should produce gamma-ray emission that could be studied by future MeV-scale space-based telescopes, such as AMEGO-X or e-ASTROGAM.

\begin{acknowledgments}

We would like to thank Kohta Murase for helpful discussions. DH is supported by the Kavli Institute for Cosmological Physics at the University of Chicago through an endowment from the Kavli Foundation and its founder Fred Kavli. DH and EP are supported by the Fermi Research Alliance, LLC under Contract No.~DE-AC02-07CH11359 with the U.S. Department of Energy, Office of Science, Office of High Energy Physics. The work of C.B.~was supported in part by NASA through the NASA Hubble Fellowship Program grant HST-HF2-51451.001-A awarded by the Space Telescope Science Institute, which is operated by the Association of Universities for Research in Astronomy, Inc., for NASA, under contract NAS5-26555. TL is supported by the European Research Council under Grant No. 742104, the Swedish National Space
Agency under contract 117/19 and by the Swedish Research Council (VR) under grants 2018-03641 and 2019-02337. This work was performed in part at the Aspen Center for Physics, which is supported by National Science Foundation grant PHY-2210452.

\end{acknowledgments}

\bibliographystyle{JHEP}
\bibliography{ref}

\providecommand{\href}[2]{#2}\begingroup\raggedright\begin{thebibliography}{10}

\bibitem{IceCube:2022der}
{\scshape IceCube} collaboration, \emph{{Evidence for neutrino emission from
  the nearby active galaxy NGC 1068}},
  \href{https://doi.org/10.1126/science.abg3395}{\emph{Science} {\bfseries 378}
  (2022) 538} [\href{https://arxiv.org/abs/2211.09972}{{\ttfamily
  2211.09972}}].

\bibitem{IceCube:2019cia}
{\scshape IceCube} collaboration, \emph{{Time-Integrated Neutrino Source
  Searches with 10 Years of IceCube Data}},
  \href{https://doi.org/10.1103/PhysRevLett.124.051103}{\emph{Phys. Rev. Lett.}
  {\bfseries 124} (2020) 051103}
  [\href{https://arxiv.org/abs/1910.08488}{{\ttfamily 1910.08488}}].

\bibitem{Fermi-LAT:2019yla}
{\scshape Fermi-LAT} collaboration, \emph{{$Fermi$ Large Area Telescope Fourth
  Source Catalog}},
  \href{https://doi.org/10.3847/1538-4365/ab6bcb}{\emph{Astrophys. J. Suppl.}
  {\bfseries 247} (2020) 33}
  [\href{https://arxiv.org/abs/1902.10045}{{\ttfamily 1902.10045}}].

\bibitem{Fermi-LAT:2019pir}
{\scshape Fermi-LAT} collaboration, \emph{{The Fourth Catalog of Active
  Galactic Nuclei Detected by the Fermi Large Area Telescope}},
  \href{https://doi.org/10.3847/1538-4357/ab791e}{\emph{Astrophys. J.}
  {\bfseries 892} (2020) 105}
  [\href{https://arxiv.org/abs/1905.10771}{{\ttfamily 1905.10771}}].

\bibitem{MAGIC:2019fvw}
{\scshape MAGIC} collaboration, \emph{{Constraints on gamma-ray and neutrino
  emission from NGC 1068 with the MAGIC telescopes}},
  \href{https://doi.org/10.3847/1538-4357/ab3a51}{\emph{Astrophys. J.}
  {\bfseries 883} (2019) 135}
  [\href{https://arxiv.org/abs/1906.10954}{{\ttfamily 1906.10954}}].

\bibitem{Murase:2022dog}
K.~Murase, \emph{{Hidden Hearts of Neutrino Active Galaxies}},
  \href{https://doi.org/10.3847/2041-8213/aca53c}{\emph{Astrophys. J. Lett.}
  {\bfseries 941} (2022) L17}
  [\href{https://arxiv.org/abs/2211.04460}{{\ttfamily 2211.04460}}].

\bibitem{IceCube:2020acn}
{\scshape IceCube} collaboration, \emph{{Characteristics of the diffuse
  astrophysical electron and tau neutrino flux with six years of IceCube high
  energy cascade data}},
  \href{https://doi.org/10.1103/PhysRevLett.125.121104}{\emph{Phys. Rev. Lett.}
  {\bfseries 125} (2020) 121104}
  [\href{https://arxiv.org/abs/2001.09520}{{\ttfamily 2001.09520}}].

\bibitem{IceCube:2021rpz}
{\scshape IceCube} collaboration, \emph{{Detection of a particle shower at the
  Glashow resonance with IceCube}},
  \href{https://doi.org/10.1038/s41586-021-03256-1}{\emph{Nature} {\bfseries
  591} (2021) 220} [\href{https://arxiv.org/abs/2110.15051}{{\ttfamily
  2110.15051}}].

\bibitem{IceCube:2013cdw}
{\scshape IceCube} collaboration, \emph{{First observation of PeV-energy
  neutrinos with IceCube}},
  \href{https://doi.org/10.1103/PhysRevLett.111.021103}{\emph{Phys. Rev. Lett.}
  {\bfseries 111} (2013) 021103}
  [\href{https://arxiv.org/abs/1304.5356}{{\ttfamily 1304.5356}}].

\bibitem{IceCube:2013low}
{\scshape IceCube} collaboration, \emph{{Evidence for High-Energy
  Extraterrestrial Neutrinos at the IceCube Detector}},
  \href{https://doi.org/10.1126/science.1242856}{\emph{Science} {\bfseries 342}
  (2013) 1242856} [\href{https://arxiv.org/abs/1311.5238}{{\ttfamily
  1311.5238}}].

\bibitem{IceCube:2014stg}
{\scshape IceCube} collaboration, \emph{{Observation of High-Energy
  Astrophysical Neutrinos in Three Years of IceCube Data}},
  \href{https://doi.org/10.1103/PhysRevLett.113.101101}{\emph{Phys. Rev. Lett.}
  {\bfseries 113} (2014) 101101}
  [\href{https://arxiv.org/abs/1405.5303}{{\ttfamily 1405.5303}}].

\bibitem{Hooper:2016jls}
D.~Hooper, \emph{{A Case for Radio Galaxies as the Sources of IceCube's
  Astrophysical Neutrino Flux}},
  \href{https://doi.org/10.1088/1475-7516/2016/09/002}{\emph{JCAP} {\bfseries
  09} (2016) 002} [\href{https://arxiv.org/abs/1605.06504}{{\ttfamily
  1605.06504}}].

\bibitem{Murase:2015xka}
K.~Murase, D.~Guetta and M.~Ahlers, \emph{{Hidden Cosmic-Ray Accelerators as an
  Origin of TeV-PeV Cosmic Neutrinos}},
  \href{https://doi.org/10.1103/PhysRevLett.116.071101}{\emph{Phys. Rev. Lett.}
  {\bfseries 116} (2016) 071101}
  [\href{https://arxiv.org/abs/1509.00805}{{\ttfamily 1509.00805}}].

\bibitem{Giacinti:2015pya}
G.~Giacinti, M.~Kachelrie\ss{}, O.~Kalashev, A.~Neronov and D.V.~Semikoz,
  \emph{{Unified model for cosmic rays above 10$^{17}$ eV and the diffuse
  gamma-ray and neutrino backgrounds}},
  \href{https://doi.org/10.1103/PhysRevD.92.083016}{\emph{Phys. Rev. D}
  {\bfseries 92} (2015) 083016}
  [\href{https://arxiv.org/abs/1507.07534}{{\ttfamily 1507.07534}}].

\bibitem{Murase:2013rfa}
K.~Murase, M.~Ahlers and B.C.~Lacki, \emph{{Testing the Hadronuclear Origin of
  PeV Neutrinos Observed with IceCube}},
  \href{https://doi.org/10.1103/PhysRevD.88.121301}{\emph{Phys. Rev. D}
  {\bfseries 88} (2013) 121301}
  [\href{https://arxiv.org/abs/1306.3417}{{\ttfamily 1306.3417}}].

\bibitem{IceCube:2023htm}
{\scshape IceCube} collaboration, \emph{{Search for correlations of high-energy
  neutrinos detected in IceCube with radio-bright AGN and gamma-ray emission
  from blazars}},  \href{https://arxiv.org/abs/2304.12675}{{\ttfamily
  2304.12675}}.

\bibitem{IceCube:2016tpw}
{\scshape IceCube} collaboration, \emph{{All-sky Search for Time-integrated
  Neutrino Emission from Astrophysical Sources with 7 yr of IceCube Data}},
  \href{https://doi.org/10.3847/1538-4357/835/2/151}{\emph{Astrophys. J.}
  {\bfseries 835} (2017) 151}
  [\href{https://arxiv.org/abs/1609.04981}{{\ttfamily 1609.04981}}].

\bibitem{IceCube:2018omy}
{\scshape IceCube} collaboration, \emph{{Constraints on minute-scale transient
  astrophysical neutrino sources}},
  \href{https://doi.org/10.1103/PhysRevLett.122.051102}{\emph{Phys. Rev. Lett.}
  {\bfseries 122} (2019) 051102}
  [\href{https://arxiv.org/abs/1807.11492}{{\ttfamily 1807.11492}}].

\bibitem{IceCube:2016ipa}
{\scshape IceCube} collaboration, \emph{{An All-Sky Search for Three Flavors of
  Neutrinos from Gamma-Ray Bursts with the IceCube Neutrino Observatory}},
  \href{https://doi.org/10.3847/0004-637X/824/2/115}{\emph{Astrophys. J.}
  {\bfseries 824} (2016) 115}
  [\href{https://arxiv.org/abs/1601.06484}{{\ttfamily 1601.06484}}].

\bibitem{Smith:2020oac}
D.~Smith, D.~Hooper and A.~Vieregg, \emph{{Revisiting AGN as the source of
  IceCube\textquoteright{}s diffuse neutrino flux}},
  \href{https://doi.org/10.1088/1475-7516/2021/03/031}{\emph{JCAP} {\bfseries
  03} (2021) 031} [\href{https://arxiv.org/abs/2007.12706}{{\ttfamily
  2007.12706}}].

\bibitem{IceCube:2016qvd}
{\scshape IceCube} collaboration, \emph{{The contribution of Fermi-2LAC blazars
  to the diffuse TeV-PeV neutrino flux}},
  \href{https://doi.org/10.3847/1538-4357/835/1/45}{\emph{Astrophys. J.}
  {\bfseries 835} (2017) 45}
  [\href{https://arxiv.org/abs/1611.03874}{{\ttfamily 1611.03874}}].

\bibitem{Hooper:2018wyk}
D.~Hooper, T.~Linden and A.~Vieregg, \emph{{Active Galactic Nuclei and the
  Origin of IceCube's Diffuse Neutrino Flux}},
  \href{https://doi.org/10.1088/1475-7516/2019/02/012}{\emph{JCAP} {\bfseries
  02} (2019) 012} [\href{https://arxiv.org/abs/1810.02823}{{\ttfamily
  1810.02823}}].

\bibitem{Khiali:2015tfa}
B.~Khiali and E.M.~de~Gouveia Dal~Pino, \emph{{High-energy neutrino emission
  from the core of low luminosity AGNs triggered by magnetic reconnection
  acceleration}}, \href{https://doi.org/10.1093/mnras/stv2337}{\emph{Mon. Not.
  Roy. Astron. Soc.} {\bfseries 455} (2016) 838}
  [\href{https://arxiv.org/abs/1506.01063}{{\ttfamily 1506.01063}}].

\bibitem{Stecker:2013fxa}
F.W.~Stecker, \emph{{PeV neutrinos observed by IceCube from cores of active
  galactic nuclei}},
  \href{https://doi.org/10.1103/PhysRevD.88.047301}{\emph{Phys. Rev. D}
  {\bfseries 88} (2013) 047301}
  [\href{https://arxiv.org/abs/1305.7404}{{\ttfamily 1305.7404}}].

\bibitem{Kimura:2014jba}
S.S.~Kimura, K.~Murase and K.~Toma, \emph{{Neutrino and Cosmic-Ray Emission and
  Cumulative Background from Radiatively Inefficient Accretion Flows in
  Low-Luminosity Active Galactic Nuclei}},
  \href{https://doi.org/10.1088/0004-637X/806/2/159}{\emph{Astrophys. J.}
  {\bfseries 806} (2015) 159}
  [\href{https://arxiv.org/abs/1411.3588}{{\ttfamily 1411.3588}}].

\bibitem{Kalashev:2014vya}
O.~Kalashev, D.~Semikoz and I.~Tkachev, \emph{{Neutrinos in IceCube from active
  galactic nuclei}}, \href{https://doi.org/10.1134/S106377611503022X}{\emph{J.
  Exp. Theor. Phys.} {\bfseries 120} (2015) 541}
  [\href{https://arxiv.org/abs/1410.8124}{{\ttfamily 1410.8124}}].

\bibitem{Murase:2015ndr}
K.~Murase, \emph{{Active Galactic Nuclei as High-Energy Neutrino Sources}},  in
  \emph{{Neutrino Astronomy}: {Current Status, Future Prospects}}, T.~Gaisser
  and A.~Karle, eds., pp.~15--31 (2017),
  \href{https://doi.org/10.1142/9789814759410_0002}{DOI}
  [\href{https://arxiv.org/abs/1511.01590}{{\ttfamily 1511.01590}}].

\bibitem{Marinucci:2015fqo}
A.~Marinucci et~al., \emph{{NuSTAR catches the unveiling nucleus of NGC 1068}},
  \href{https://doi.org/10.1093/mnrasl/slv178}{\emph{Mon. Not. Roy. Astron.
  Soc.} {\bfseries 456} (2016) L94}
  [\href{https://arxiv.org/abs/1511.03503}{{\ttfamily 1511.03503}}].

\bibitem{Bauer:2014rla}
F.E.~Bauer et~al., \emph{{NuSTAR Spectroscopy of Multi-Component X-ray
  Reflection from NGC 1068}},
  \href{https://doi.org/10.1088/0004-637X/812/2/116}{\emph{Astrophys. J.}
  {\bfseries 812} (2015) 116}
  [\href{https://arxiv.org/abs/1411.0670}{{\ttfamily 1411.0670}}].

\bibitem{AMEGO:2019gny}
{\scshape AMEGO} collaboration, \emph{{All-sky Medium Energy Gamma-ray
  Observatory: Exploring the Extreme Multimessenger Universe}},
  \href{https://arxiv.org/abs/1907.07558}{{\ttfamily 1907.07558}}.

\bibitem{e-ASTROGAM:2016bph}
{\scshape e-ASTROGAM} collaboration, \emph{{The e-ASTROGAM mission}},
  \href{https://doi.org/10.1007/s10686-017-9533-6}{\emph{Exper. Astron.}
  {\bfseries 44} (2017) 25} [\href{https://arxiv.org/abs/1611.02232}{{\ttfamily
  1611.02232}}].

\bibitem{Murase:2019vdl}
K.~Murase, S.S.~Kimura and P.~Meszaros, \emph{{Hidden Cores of Active Galactic
  Nuclei as the Origin of Medium-Energy Neutrinos: Critical Tests with the MeV
  Gamma-Ray Connection}},
  \href{https://doi.org/10.1103/PhysRevLett.125.011101}{\emph{Phys. Rev. Lett.}
  {\bfseries 125} (2020) 011101}
  [\href{https://arxiv.org/abs/1904.04226}{{\ttfamily 1904.04226}}].

\bibitem{Dermer:1995ju}
C.D.~Dermer, J.A.~Miller and H.~Li, \emph{{Stochastic particle acceleration
  near accreting black holes}},
  \href{https://doi.org/10.1086/176631}{\emph{Astrophys. J.} {\bfseries 456}
  (1996) 106} [\href{https://arxiv.org/abs/astro-ph/9508069}{{\ttfamily
  astro-ph/9508069}}].

\bibitem{Dermer:2014vaa}
C.D.~Dermer, K.~Murase and Y.~Inoue, \emph{{Photopion Production in Black-Hole
  Jets and Flat-Spectrum Radio Quasars as PeV Neutrino Sources}},
  \href{https://doi.org/10.1016/j.jheap.2014.09.001}{\emph{JHEAp} {\bfseries
  3-4} (2014) 29} [\href{https://arxiv.org/abs/1406.2633}{{\ttfamily
  1406.2633}}].

\bibitem{Murase:2011cx}
K.~Murase, K.~Asano, T.~Terasawa and P.~Meszaros, \emph{{The Role of Stochastic
  Acceleration in the Prompt Emission of Gamma-Ray Bursts: Application to
  Hadronic Injection}},
  \href{https://doi.org/10.1088/0004-637X/746/2/164}{\emph{Astrophys. J.}
  {\bfseries 746} (2012) 164}
  [\href{https://arxiv.org/abs/1107.5575}{{\ttfamily 1107.5575}}].

\bibitem{Stawarz:2008sp}
L.~Stawarz and V.~Petrosian, \emph{{On the Momentum Diffusion of Radiating
  Ultrarelativistic Electrons in a Turbulent Magnetic Field}},
  \href{https://doi.org/10.1086/588813}{\emph{Astrophys. J.} {\bfseries 681}
  (2008) 1725} [\href{https://arxiv.org/abs/0803.0989}{{\ttfamily 0803.0989}}].

\bibitem{Musulmanbekov:2003wy}
G.~Musulmanbekov, \emph{{Total cross-section, inelasticity and multiplicity
  distributions in proton proton collisions}},
  \href{https://doi.org/10.1134/1.1644012}{\emph{Phys. Atom. Nucl.} {\bfseries
  67} (2004) 90} [\href{https://arxiv.org/abs/hep-ph/0304304}{{\ttfamily
  hep-ph/0304304}}].

\bibitem{Kamae:2006bf}
T.~Kamae, N.~Karlsson, T.~Mizuno, T.~Abe and T.~Koi, \emph{{Parameterization of
  Gamma, e+/- and Neutrino Spectra Produced by p-p Interaction in Astronomical
  Environment}}, \href{https://doi.org/10.1086/513602}{\emph{Astrophys. J.}
  {\bfseries 647} (2006) 692}
  [\href{https://arxiv.org/abs/astro-ph/0605581}{{\ttfamily
  astro-ph/0605581}}].

\bibitem{BeckerTjus:2014uyv}
J.~Becker~Tjus, B.~Eichmann, F.~Halzen, A.~Kheirandish and S.M.~Saba,
  \emph{{High-energy neutrinos from radio galaxies}},
  \href{https://doi.org/10.1103/PhysRevD.89.123005}{\emph{Phys. Rev. D}
  {\bfseries 89} (2014) 123005}
  [\href{https://arxiv.org/abs/1406.0506}{{\ttfamily 1406.0506}}].

\bibitem{Mannheim:1994sv}
K.~Mannheim and R.~Schlickeiser, \emph{{Interactions of Cosmic Ray Nuclei}},
  {\emph{Astron. Astrophys.} {\bfseries 286} (1994) 983}
  [\href{https://arxiv.org/abs/astro-ph/9402042}{{\ttfamily
  astro-ph/9402042}}].

\bibitem{Breit:1934zz}
G.~Breit and J.A.~Wheeler, \emph{{Collision of two light quanta}},
  \href{https://doi.org/10.1103/PhysRev.46.1087}{\emph{Phys. Rev.} {\bfseries
  46} (1934) 1087}.

\bibitem{Gould:1967zzb}
R.J.~Gould and G.P.~Schreder, \emph{{Pair Production in Photon-Photon
  Collisions}}, \href{https://doi.org/10.1103/PhysRev.155.1404}{\emph{Phys.
  Rev.} {\bfseries 155} (1967) 1404}.

\bibitem{1983Afz....19..323A}
F.A.~{Aharonian}, A.M.~{Atoian} and A.M.~{Nagapetian}, \emph{{Photoproduction
  of electron-positron pairs in compact X-ray sources}}, {\emph{Astrofizika}
  {\bfseries 19} (1983) 323}.

\bibitem{Fang:2022trf}
K.~Fang, J.S.~Gallagher and F.~Halzen, \emph{{The TeV Diffuse Cosmic Neutrino
  Spectrum and the Nature of Astrophysical Neutrino Sources}},
  \href{https://doi.org/10.3847/1538-4357/ac7649}{\emph{Astrophys. J.}
  {\bfseries 933} (2022) 190}
  [\href{https://arxiv.org/abs/2205.03740}{{\ttfamily 2205.03740}}].

\bibitem{1981Ap&SS..79..321A}
F.A.~{Aharonian} and A.M.~{Atoyan}, \emph{{Compton Scattering of Relativistic
  Electrons in Compact X-Ray Sources}},
  \href{https://doi.org/10.1007/BF00649428}{\emph{Astrophysics and Space
  Science} {\bfseries 79} (1981) 321}.

\bibitem{Fornengo:2011iq}
N.~Fornengo, R.A.~Lineros, M.~Regis and M.~Taoso, \emph{{Galactic synchrotron
  emission from WIMPs at radio frequencies}},
  \href{https://doi.org/10.1088/1475-7516/2012/01/005}{\emph{JCAP} {\bfseries
  01} (2012) 005} [\href{https://arxiv.org/abs/1110.4337}{{\ttfamily
  1110.4337}}].

\bibitem{ricci2017bat}
C.~Ricci, B.~Trakhtenbrot, M.J.~Koss, Y.~Ueda, I.~Del~Vecchio, E.~Treister
  et~al., \emph{Bat agn spectroscopic survey. v. x-ray properties of the
  swift/bat 70-month agn catalog}, {\emph{The Astrophysical Journal Supplement
  Series} {\bfseries 233} (2017) 17}.

\bibitem{2017ICRC...35..824W}
M.~{Wood}, R.~{Caputo}, E.~{Charles}, M.~{Di Mauro}, J.~{Magill},
  J.S.~{Perkins} et~al., \emph{{Fermipy: An open-source Python package for
  analysis of Fermi-LAT Data}},  in \emph{35th International Cosmic Ray
  Conference (ICRC2017)}, vol.~301 of \emph{International Cosmic Ray
  Conference}, p.~824, July, 2017,
  \href{https://doi.org/10.22323/1.301.0824}{DOI}
  [\href{https://arxiv.org/abs/1707.09551}{{\ttfamily 1707.09551}}].

\bibitem{Peretti:2023crf}
E.~Peretti, G.~Peron, F.~Tombesi, A.~Lamastra, M.~Ahlers and F.G.~Saturni,
  \emph{{Gamma-ray emission from the Seyfert galaxy NGC 4151 and multimessenger
  implications for ultra-fast outflows}},
  \href{https://arxiv.org/abs/2303.03298}{{\ttfamily 2303.03298}}.

\bibitem{Neronov:2023aks}
A.~Neronov, D.~Savchenko and D.V.~Semikoz, \emph{{Neutrino signal from Seyfert
  galaxies}},  \href{https://arxiv.org/abs/2306.09018}{{\ttfamily 2306.09018}}.

\bibitem{ackermann2012search}
M.~Ackermann, M.~Ajello, A.~Allafort, L.~Baldini, J.~Ballet, G.~Barbiellini
  et~al., \emph{Search for gamma-ray emission from x-ray-selected seyfert
  galaxies with fermi-lat}, {\emph{The Astrophysical Journal} {\bfseries 747}
  (2012) 104}.

\bibitem{panessa2006x}
F.~Panessa, L.~Bassani, M.~Cappi, M.~Dadina, X.~Barcons, F.J.~Carrera et~al.,
  \emph{On the x-ray, optical emission line and black hole mass properties of
  local seyfert galaxies}, {\emph{Astronomy \& Astrophysics} {\bfseries 455}
  (2006) 173}.

\bibitem{perola1994x}
G.~Perola and L.~Piro, \emph{X-ray reprocessing and uv continuum in agn: The
  case of ngc 4151}, {\emph{Astronomy and Astrophysics (ISSN 0004-6361), vol.
  281, no. 1, p. 7-14} {\bfseries 281} (1994) 7}.

\bibitem{bentz2015agn}
M.C.~Bentz and S.~Katz, \emph{The agn black hole mass database},
  {\emph{Publications of the Astronomical Society of the Pacific} {\bfseries
  127} (2015) 67}.

\bibitem{yuan2020cepheid}
W.~Yuan, M.M.~Fausnaugh, S.L.~Hoffmann, L.M.~Macri, B.M.~Peterson, A.G.~Riess
  et~al., \emph{The cepheid distance to the seyfert 1 galaxy ngc 4151},
  {\emph{The Astrophysical Journal} {\bfseries 902} (2020) 26}.

\bibitem{masini2018obscured}
A.~Masini, \emph{Obscured and compton-thick agn in nustar hard x-ray surveys},
  .

\end{thebibliography}\endgroup

\end{document}